# Incorporating ESO into Deep Koopman Operator Modelling for Control of Autonomous Vehicles

Hao Chen, Chen Lv, *Senior Member, IEEE*

*Abstract*—Koopman operator theory is a kind of data-driven modelling approach that accurately captures the nonlinearities of mechatronic systems such as vehicles against physics-based methods. However, the infinite-dimensional Koopman operator is impossible to implement in real-world applications. To approximate the infinite-dimensional Koopman operator through collection dataset rather than manual trial and error, we adopt deep neural networks (DNNs) to extract basis functions by offline training and map the nonlinearities of vehicle planar dynamics into a linear form in the lifted space. Besides, the effects of the dimensions of basis functions on the model accuracy are explored. Further, the extended state observer (ESO) is introduced to online estimate the total disturbance in the lifted space and compensate for the modelling errors and residuals of the learned deep Koopman operator (DK) while also improving its generalization. Then, the proposed model is applied to predict vehicle states within prediction horizons and later formulates the constrained finite-time optimization problem of model predictive control (MPC), i.e., ESO-DKMPC. In terms of the trajectory tracking of autonomous vehicles, the ESO-DKMPC generates the wheel steering angle to govern lateral motions based on the decoupling control structure. The various conditions under the double-lane change scenarios are built on the CarSim/Simulink co-simulation platform, and extensive comparisons are conducted with the linear MPC (LMPC) and nonlinear MPC (NMPC) informed by the physics-based model. The results indicate that the proposed ESO-DKMPC has better tracking performance and moderate efficacy both within linear and nonlinear regions.

## I. INTRODUCTION

### A. Motivation and Related wor

SYSTEM modelling serves as the foundation for control and optimization, while model accuracy is the key to the design and performance of subsequent controllers [1]. In the case of nonlinear mechatronic systems, such as vehicles, it is difficult to identify nonlinear behaviors using a physics-based model. Some procedures can only be approximated or fitted using empirical models, but calibrating the hyperparameters within these empirical models is a challenging task [2]–[5]. In turn, this poses difficulties for the formulation of control laws. The methodology of data-driven modelling offers a comprehensive guide to addressing these concerns. The data-driven approach does not depend on the internal physics of control plants. Instead, it utilizes input and output data to efficiently capture the characteristics of nonlinear components and enhance the accuracy of the system model [6]–[9].

Unlike the local linearization [10], [11] or backstepping [12], [13] commonly used for nonlinear systems, Koopman operator lifts the original nonlinear state space into an infinite-dimensional linear state space through data [14]–[18]. The mapping from the original low-dimensional space to the infinite-dimensional lifted space enforces the nonlinearities to be linear behaviors, resulting in a linear representation of the nonlinear system. Therefore, the methodologies that have been established for the analysis and synthesis of linear systems have the potential for further extension. In addition, the original space can be embedded into the lifted space, i.e., subspace, and retrieved from the lifted space without any loss of accuracy. Petar Bevanda et al. have proven that Koopman representation has marginally better performance than other alternative data-driven modelling techniques, particularly in the context of nonlinear systems [19]. However, the infinite-dimensional Koopman operator is unattainable in applications. Thus, the suitable choice of finite-dimensional basis functions is considered to approximate the Koopman operator.

Deep learning method has been demonstrated in comparison to the two prevailing mainstream approaches, namely dynamic mode decomposition (DMD) [20]–[22] and extended dynamic mode decomposition (EDMD) [21], [23], [24]. DMD and EDMD make a choice of basis functions based on a priori knowledge or by a non-trivial process of trial and error, thus restricting the generalization. Whereas the deep learning approach only builds the deep neural network (DNN), which is a kind of nonlinear mapping rather than DMD as linear mapping, and helps the Koopman operator to capture nonlinearities of the original space and evolve with linear form in the lifted space. Moreover, it can extract basis functions through an automated procedure following training [25]. Yiqiang Han et al. employed the DNN for the data-driven identification of the appropriate basis functions to exploit the power of data. The OpenAI Gym environment is used for data generation and training of DNN to learn the basis functions of Koopman operator. This particular controller is verified to be sufficient for the reinforcement learning algorithm [26]. Yongqian Xiao et al. introduced a DNN-based vehicle modelling approach with an interpretable Koopman operator. A model predictive controller with the learned Koopman model for velocity profile tracking of autonomous vehicles shows better performance and higher computational efficiency than other traditional and advanced modeling methods [27].

On the other hand, the finite-dimensional approximation for the Koopman operator introduces modelling errors, and residuals are inherent during offline network training. Moreover,

Hao Chen, and Chen Lv are with the School of Mechanical and Aerospace Engineering, Nanyang Technological University, 639798, Singapore, e-mails: chen.h@ntu.edu.sg, lyuchen@ntu.edu.sg.
Corresponding author: C. Lv

vehicles exhibit time-varying dynamics when operating in complicated environments. If the real-world scenario extends beyond the feature space identified by the training dataset, the learned Koopman operator may result in inaccuracies, even faults. The extended state observer (ESO) augments the unknown components of the system, including modelling errors and uncertainties, as its extended state while replacing the linear feedback functions with a nonlinear form. This renders the observer not reliant on the precise physics model but on the necessary measurements, with promising convergency [28]. Jesu´s Guerrero et al. designed adaptive ESO to observe unknown external disturbances and parametric uncertainties of underwater vehicles in finite time. With the adaptive laws to update the gains of ESO, the observer can reject bounded time-varying disturbances even if the upper bound of the perturbation is not known. Additionally, the stability of the observer scheme is proven using Lyapunov's arguments [29]. Chao Ren et al. proposed a comprehensive Koopman operator-based robust data-driven control framework for wheeled mobile robots. The EDMD approach is applied to obtain the Koopman operator, and the modelling errors are online estimated by ESO and compensated in the control signal from the sliding mode controller. Experimental tests show robustness against disturbances [30].

### B. Contribution

Motivated by the aforementioned above, this paper adopts DNNs to approximate the infinite-dimensional Koopman operator, i.e., the deep Koopman operator, to feature the vehicle dynamics based on the collection dataset. To reduce the modelling errors and residuals, ESO is specifically designed to online estimate the total disturbances to improve the accuracy of the deep Koopman operator model. Meanwhile, the decoupling control structure is utilized in order to simplify the trajectory tracking of autonomous vehicles. The ESO-DKMPC is developed to govern the vehicle's lateral motions. The main contributions of this paper are listed as follows:

- ESO is designed to online estimate the modelling errors and residuals, afterward as a complement to the learned deep Koopman operator to enhance its model fidelity and generalization. This is helpful for analyzing vehicle planar dynamics in the lifted space.
- In terms of the vehicle's lateral motion regulation, by incorporating ESO, the deep Koopman operator model acts as the predictive model to formulate the constrained finite-time optimization problem of the model predictive control scheme in the lifted space, referred to as ESO-DKMPC.
- The effects of the dimensions of basis functions on the model fidelity are investigated, and various conditions are built on the CarSim/Simulink co-simulation platform to conduct rigorous comparisons with other methods.

### C. Paper organization

The remaining paper is organized as follows: Section II provides a brief overview of the methodologies of deep learning Koopman operator, ESO, and the ESO-DKMPC. The framework of ESO-DKMPC for trajectory tracking of autonomous vehicles, including the decoupling control structure, the deep Koopman operator model with ESO regarding vehicle dynamics, and the corresponding constrained finite-time optimization problem, are formulated in Section III. Moreover, the dimensions of the basis functions are also discussed therein. Section IV validates the accuracy and robustness of the deep Koopman model, which incorporates an online estimated total disturbance by ESO. In addition, the comparison studies of four typical methods of tracking performance are conducted under various double-lane change scenarios on the CarSim/Simulink co-simulation platform. Section V concludes the paper.

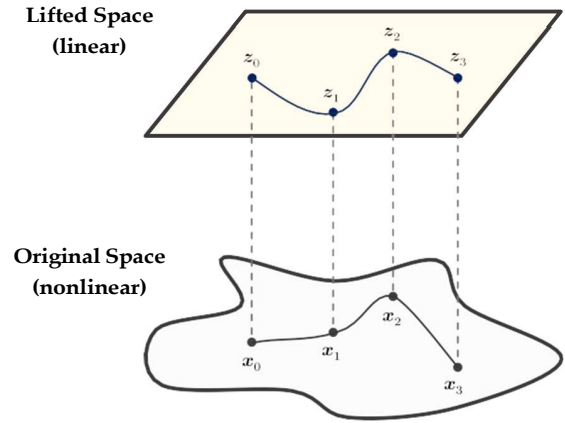

Fig. 1. Koopman operator $\kappa$ lifting.

## II. METHODOLOGY

### A. Deep learning Koopman operator

A general discrete-time nonlinear system can be defined as:

$$\boldsymbol{x}_{k+1} = \boldsymbol{f}(\boldsymbol{x}_k, \boldsymbol{u}_k) \quad (1)$$

where $\boldsymbol{x}_k \in \mathbf{R}^n$ is the state vector at time $k$, $\boldsymbol{u}_k \in \mathbf{R}^m$ is control input vector, and $\boldsymbol{f} : \mathbf{R}^{n+m} \to \mathbf{R}^n$ is a smooth vector field of $\mathbf{C}^\infty$.

Koopman operator, denoted by $\kappa$, is an infinite linear operator defined on the space of embedding functions $\boldsymbol{g}$ as:

$$\kappa \boldsymbol{g}(\boldsymbol{x}_k, \boldsymbol{u}_k) = \boldsymbol{g}(\boldsymbol{x}_{k+1}, \boldsymbol{u}_{k+1}) = \boldsymbol{g}(\boldsymbol{f}(\boldsymbol{x}_k, \boldsymbol{u}_k), \boldsymbol{u}_{k+1}) \quad (2)$$

where $\boldsymbol{g} : \mathbf{R}^{m+n} \to \mathbf{R}^d$ is a smooth vector field of $\mathbf{C}^\infty$.

Therefore, the original state space has been lifted into the designated embedding space by $\boldsymbol{g}$. This mapping, characterized by the Koopman operator $\kappa$ (Fig. 1), allows the nonlinear dynamics inherent in the original state space to be linear.

The embedding function $\boldsymbol{g}$ can be further divided into two components:

$$\boldsymbol{g}(\boldsymbol{x}_k, \boldsymbol{u}_k) = \begin{bmatrix} \boldsymbol{g}_x(\boldsymbol{x}_k) \\ \boldsymbol{u}_k \end{bmatrix} \quad (3)$$

where $\boldsymbol{g}_x : \mathbf{R}^n \to \mathbf{R}^{d-m}$.

However, the infinite-dimensional $\kappa$ is impossible to store and compute in real applications but can be approximated by a finite-dimensional representation. Then, without loss of generality, (2) is simplified within finite dimension space as:

$$\begin{bmatrix} g_x(x_{k+1}) \\ u_{k+1} \end{bmatrix} = \begin{bmatrix} K_{xx} & K_{xu} \\ K_{ux} & K_{uu} \end{bmatrix} \begin{bmatrix} g_x(x_k) \\ u_k \end{bmatrix} \quad (4)$$

where $K \in \mathbf{R}^{d \times d}$ is the finite-dimensional approximation of $\kappa$ in the lifted space.

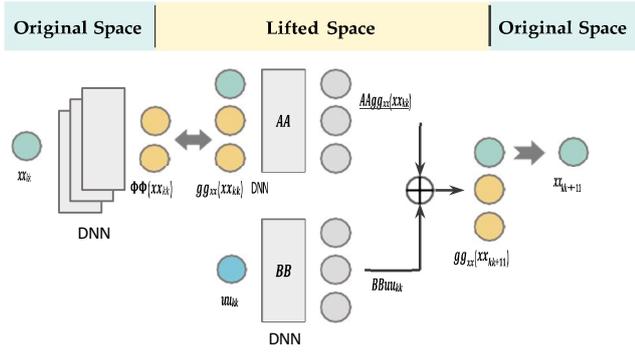

Fig. 2. The framework of deep Koopman operator.

From (4), we denote $K_{xx} = A$, $K_{xu} = B$, and the time-invariant state evolution in the lifted space can be written as:

$$g_x(x_{k+1}) = A g_x(x_k) + B u_k \quad (5)$$

where $A \in \mathbf{R}^{(d-m) \times (d-m)}$ and $B \in \mathbf{R}^{(d-m) \times m}$ are the finite-dimensional approximation of $\kappa$ in terms of states. In terms of the input evolution, i.e., $u_{k+1} = K_{ux} g_x(x_k) + K_{uu} u_k$, $u_{k+1}$ is free to be chosen. Therefore, $K_{ux}$ and $K_{uu}$ are time-varying, since they are not concluded until $u_{k+1}$ is specified.

Further, to reconstruct the original state space from the lifted space at time $k$, $g_x$ is defined as:

$$g_x(x_k) := z_k = \begin{bmatrix} x_k \\ \Phi_k(x_k) \end{bmatrix} \quad (6)$$

where

$\Phi_k(x_k) = [\varphi_{1,k}(x_k), \varphi_{2,k}(x_k), \ldots, \varphi_{d-m-n,k}(x_k)]^T \in \mathbf{R}^{d-m-n}$

is the basis in the lifted space.

From (6), we have:

$$x_k = C_P z_k \quad (7)$$

where $C_P = [I_n, 0] \in \mathbf{R}^{n \times (d-m)}$ is the projection matrix from the lifted space to the state space.

Consequently, the state space is encompassed within the lifted space, enabling its faithful reconstruction without any loss of accuracy through the $g_x$ projection rather than these approximations in [21].

The aforementioned analysis reveals that the choice of $\Phi$ determines the approximation accuracy of $K$ in (4). Whereas DNN is regarded as a set of nonlinear basis functions to approximate complex relationships. Therefore, the basis $\Phi$ can be directly learned by training DNN from data pairs, which enforces linear dynamics in the lifted space (Fig. 2). Its number of hidden layers and units are manually tuning parameters to balance accuracy and complexity. Besides, two single-hidden layers, without activation and bias, are employed as the linear transformation to feature $A$ and $B$ in the lifted space (Fig. 2), respectively. This ensures the continuous derivatives of the loss function in training, thus enabling efficient error backpropagation and a quick convergence of DNN.

The loss function $L$ is defined to learn the weights and bias of DNN in (8):

$$L = L_1 + L_2 \quad (8)$$

where $L_1$ is the total prediction loss of a data sequence:

$$L_1(\theta_\Phi; \theta_A; \theta_B) = \sum_{i=1}^{p-1} \|x_{k+i} - \hat{x}_{k+i}\|_2^2 \quad (9)$$

where $p$ is the sequence length, $\theta_\Phi$, $\theta_A$, and $\theta_B$ are the learning parameters of DNNs, respectively, and $\hat{x}_{k+i}(i = 1, 2, \ldots p-1)$ is the $i$-step forward prediction from the lifted space by the deep Koopman operator, denoted as:

$$\hat{x}_{k+i} = C_P [A(\theta_A) z_{k+i-1} + B(\theta_B) u_{k+i-1}] \quad (10)$$

and $L_2$ is the reconstruction loss:

$$L_2(\theta_\Phi; \theta_{\Phi^{-1}}) = \sum_{i=1}^{p} \|x_k - \Phi^{-1}(\Phi(x_k))\|_2^2 \quad (11)$$

where $\Phi^{-1}$ is the decoder net of $\Phi$. This intricate interplay between the encoder and decoder networks enforces the encoded representation that encompasses the essential characteristics of the input data. As a result, the decoder can reconstruct the input based on the encoded features.

### B. Extended state observer

ESO augments the total disturbance of system, including unmodelled dynamics and external disturbances, as a new state variable and later estimates all the states of the extended system. Notably, ESO estimates the external disturbances and unmodelled plant dynamics using the input-output data rather than the system and input matrices. This approach enables a more comprehensive estimation framework.

The deep Koopman operator has inevitable modelling errors. Therefore, the evolution dynamics in the lifted space are adherent to:

$$z_{k+1} = A_\theta z_k + B_\theta u_k + w_k \quad (12)$$

where $z_k$ is defined in (6) with the learned basis $\Phi$, $A_\theta$ and $B_\theta$ are the learned system matrix and input matrix, respectively, and $w_k \in \mathbf{R}^{d-m}$ is the total modelling error or disturbance in the lifted space.

Based on the dynamics in the lifted space of (12), the following ESO is defined as:

$$\begin{aligned} \hat{z}_{k+1} &= A_\theta \hat{z}_k + B_\theta u_k + [w_k - \beta_1 g_1(\tilde{z}_k)] \\ w_{k+1} &= w_k - \beta_2 g_2(\tilde{z}_k) \end{aligned} \quad (13)$$

where $\hat{z}_k$ and $w_k$ are the estimated results of state vector and total disturbance in the lifted space, respectively, and $\tilde{z}_k$ is the estimation error between $z_k$ and $\hat{z}_k$ as:

$$\tilde{z}_k = z_k - \hat{z}_k \quad (14)$$

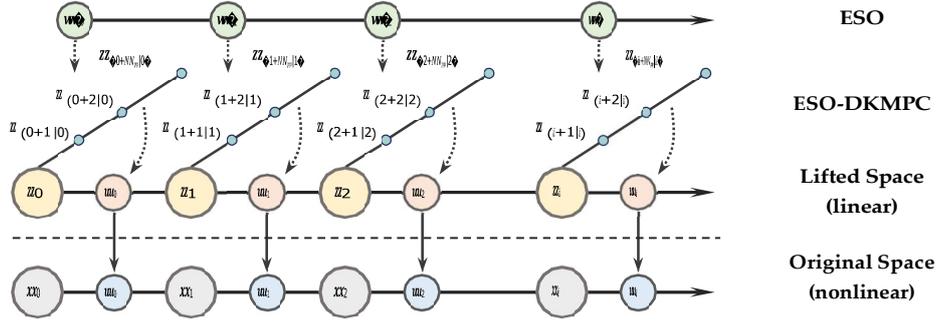

Fig. 3. Incorporating ESO into model predictive control in the lifted space.

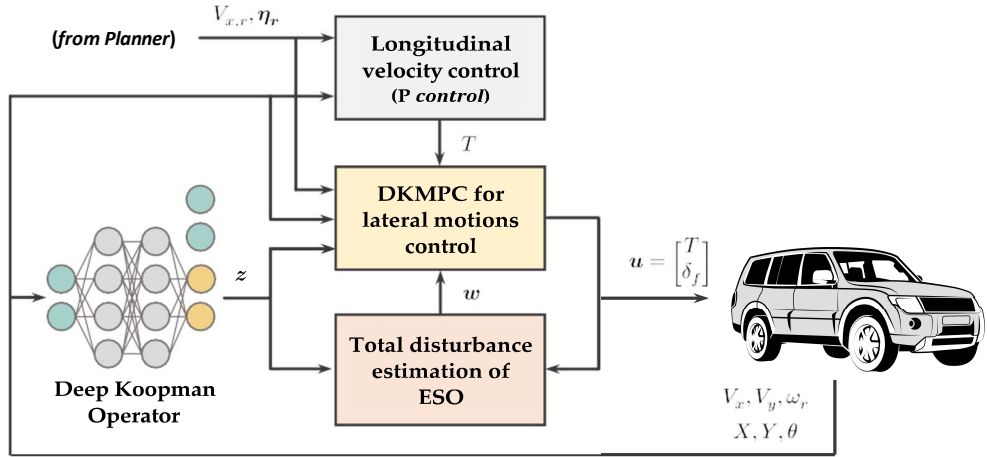

Fig. 4. The framework of ESO-DKMPC for trajectory tracking of autonomous vehicles.

$g_1$ and $g_2$ are differentiable and continuous functions around the original with the scaling factors $\beta_1$ and $\beta_2$.

*Remark 1:* By choosing appropriate differentiable and continuous functions, as well as scaling factors, (13) has been proven its convergency and consistency with (12) while providing a reliable estimation of $w_k$ at each time step (see Appendix A). To be brevity, we choose $g_1(z_k) = g_2(z_k) = z_k$ in a linear form in this work.

### C. ESO-Deep Koopman operator-informed MPC (ESO-DKMPC)

The framework of incorporating ESO into MPC in the lifted space is illustrated in Fig. 3. The total disturbance or modelling error is estimated online by ESO at each time step, and then involved in the predictive model in the lifted space to improve the model fidelity and generate the control sequence. Moreover, the updated total disturbance remains unchanged during prediction horizons.

Therefore, based on (5)∼(7) and (12), we formulate a new optimal problem of MPC in the corresponding lifted space that also integrates disturbance estimation from ESO, that is, ESO-DKMPC, as:

$$\min_{U_k} \sum_{i=1}^{N_p} \|\psi(C_P \hat{z}_{k+i|k}) - \eta_{r,k+i|k}\|_Q^2 + \sum_{i=0}^{N_c-1} \|u_{k+i|k}\|_R^2$$
$$+ \sum_{i=0}^{N_c-1} \|\Delta u_{k+i|k}\|_P^2$$

$$\hat{z}_{k+i+1|k} = A_\theta \hat{z}_{k+i|k} + B_\theta u_{k+i|k} + w_k \quad \forall i \in [0, N_c-1]$$
$$\hat{z}_{k+i+1|k} = A_\theta \hat{z}_{k+i|k} + B_\theta u_{k+N_c-1|k} + w_k \quad \forall i \in [N_c, N_p-1]$$
$$\Delta u_{k+i|k} = u_{k+i|k} - u_{k+i-1|k} \quad \forall i \in [0, N_c-1]$$
$$h_x(\hat{z}_{k+i|k}) \leq 0 \quad \forall i \in [1, N_p]$$
$$h_u(u_{k+i|k}) \leq 0 \quad \forall i \in [0, N_c-1]$$
$$\hat{z}_{k|k} = z_k, u_{k-1|k} = u_{k-1}$$
(15)

where $U_k = [u_{k|k}, u_{k+1|k}, \ldots u_{k+N_c-1|k}]^T$ is the optimal control sequence, $\psi$ is the output function in the original space, $\eta_r$ is the corresponding reference during prediction horizons, $Q$ and $P$ are semi-positive definite weight matrices for states and control increments, respectively, $R$ is positive definite weight matrix for control efforts, $N_c$ and $N_p$ are steps of control and prediction horizon, respectively, $h_x$ and $h_u$ are vector functions in terms of the constraints of $\boldsymbol{x}$ and $\boldsymbol{u}$, respectively.

*Remark 2:* Because we decouple $\boldsymbol{g}_x(\boldsymbol{x})$ and $\boldsymbol{u}$ from $\boldsymbol{g}$ while maintaining $\boldsymbol{u}$ a linear form in (3), the optimal control sequence of (15) is equivalent to that of in the original space, and the first element of $\boldsymbol{U}_k$ can be applied straight to the control plant.

## III. ESO-DKMPC FOR TRAJECTORY TRACKING OF AUTONOMOUS VEHICLES

Fig. 4 depicts the scheme of ESO-DKMPC for trajectory tracking of autonomous vehicles. We employ the decoupling structure to control the longitudinal velocity and lateral motions because it is amenable to practical implementations. Concretely, the proportional controller (P control) is designed to track the desired longitudinal velocity profile $V_{x,r}$ from the higher-level planner and obtain the total torque $T$. In terms of lateral motions, through using a learned DNN-informed Koopman operator, the longitudinal velocity $V_x$, lateral velocity $V_y$, and yaw rate $\omega_r$ of the vehicle are initially mapped into the lifted space. Meanwhile, ESO online estimates the total disturbance $\boldsymbol{w}$ in the lifted space, later involved in the predictive model. The current pose signals, i.e., positions $X$ and $Y$, yaw angle $\theta$ in the global frame, are measured from sensors [31], [32]. Consequently, concerning the total torque $T$ and the reference pose $\boldsymbol{\eta}_r$ also from the planner, the optimal control problem, ESO-DKMPC, as defined in (15), is formulated and solved to generate the control command.

### A. Vehicle Planar Dynamics

The four-wheel vehicle planar dynamics model is demonstrated in Fig. 5. The Newton-Euler equations of planar motion for the vehicle at time $k$ are [33]:

$$\begin{cases} V_{x,k+1} = V_{x,k} + T_s V_{y,k} \omega_{r,k} \\ \quad + \dfrac{T_s}{m}(F_{x3,k} + F_{x4,k}) + \dfrac{T_s}{m}(F_{x1,k} + F_{x2,k})\cos\delta_{f,k} \\ \quad - \dfrac{T_s}{m}(F_{y1,k} + F_{y2,k})\sin\delta_{f,k} \\ V_{y,k+1} = V_{y,k} - T_s V_{x,k}\omega_{r,k} \\ \quad + \dfrac{T_s}{m}(F_{y3,k} + F_{y4,k}) + \dfrac{T_s}{m}(F_{x1,k} + F_{x2,k})\sin\delta_{f,k} \\ \quad + \dfrac{T_s}{m}(F_{y1,k} + F_{y2,k})\cos\delta_{f,k} \\ \omega_{r,k+1} = \omega_{r,k} - T_s \dfrac{l_r}{I_z}(F_{y3,k} + F_{y4,k}) \\ \quad - T_s \dfrac{w_B}{2I_z}[(F_{x1,k}\cos\delta_{f,k} - F_{y1,k}\sin\delta_{f,k}) + F_{x3,k}] \\ \quad + T_s \dfrac{w_B}{2I_z}[(F_{x2,k}\cos\delta_{f,k} - F_{y2,k}\sin\delta_{f,k}) + F_{x4,k}] \\ \quad + T_s \dfrac{l_f}{I_z}(F_{x1,k}\sin\delta_{f,k} + F_{y1,k}\cos\delta_{f,k}) \\ \quad + T_s \dfrac{l_f}{I_z}(F_{x2,k}\sin\delta_{f,k} + F_{y2,k}\cos\delta_{f,k}) \end{cases} \tag{16}$$

where, $T_s$ is the discrete time step, $V_{x,k}$, $V_{y,k}$, and $\omega_{r,k}$ are the longitudinal velocity, lateral velocity, and yaw rate of the vehicle, respectively, $F_{xi,k}(i = 1, 2, 3, 4)$ is the longitudinal tire force, $F_{yi,k}(i = 1, 2, 3, 4)$ is the lateral tire force, $\delta_{f,k}$ is the steering angle of the front wheel, $m$ is the vehicle mass, $I_z$ is the mass moment, $l_f$ is the distance from the front axle to c.g., $l_r$ is the distance from the rear axle to c.g., and $w_B$ is the track width.

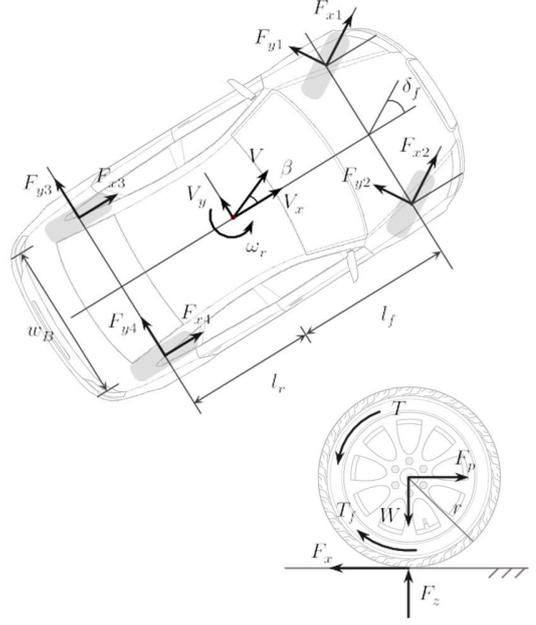

Fig. 5. The four-wheel vehicle planar dynamics model.

The longitudinal and lateral tire forces of the $i$-th ($i = 1, 2, 3, 4$) driven wheel are characterized using the empirical model [34], [35] as:

$$\begin{aligned} F_{xi,k} &= F_{xi,k}(T_{i,k}, \omega_{i,k}, F_{zi,k}, \mu) \\ F_{yi,k} &= F_{yi,k}(\alpha_{i,k}, F_{zi,k}, \mu) \end{aligned} \tag{17}$$

where $\omega_{i,k}$ is the rotational velocity, $T_{i,k}$ is the motor torque, $\alpha_{i,k}$ is the tire sideslip angle related to velocities, yaw dynamics, steering angle, and vehicle specifications [26], $F_{zi,k}$ is the normal force equal to the tire vertical load $W_{i,k}$, and $\mu$ is the road friction coefficient.

The driving torque and the steering angle are the control inputs, and the longitudinal velocity, lateral velocity, and yaw rate are the states of the vehicle planar dynamics model:

$$\begin{aligned} \boldsymbol{u}_k &= [T_k, \delta_{f,k}]^T \\ \boldsymbol{x}_k &= [V_{x,k}, V_{y,k}, \omega_{r,k}]^T \end{aligned} \tag{18}$$

where the driving torque $T_k = T_{1,k} + T_{2,k} + T_{3,k} + T_{4,k}$ is the summation of motor torque of each wheel.

Based on (14)$\sim$(16), (13) of discrete-time form is denoted as:

$$\begin{bmatrix} V_{x,k+1} \\ V_{y,k+1} \\ \omega_{r,k+1} \end{bmatrix} = \boldsymbol{f}\left(\begin{bmatrix} V_{x,k} \\ V_{y,k} \\ \omega_{r,k} \end{bmatrix}, \begin{bmatrix} T_k \\ \delta_{f,k} \end{bmatrix}\right) \tag{19}$$

where $\boldsymbol{f}$ maps the nonlinear evolution dynamics of the vehicle in a planar motion consistent with that of (1).

TABLE I
STATISTIC RESULTS OF ONE-STEP PREDICTION ERRORS WITH RESPECT TO DIFFERENT DIMENSIONS

| | $E_x$ (m/s) | $E_y$ (m/s) | $E_r$ (rad/s) |
|---|---|---|---|
| | Max / Avg / RMSE | Max / Avg / RMSE | Max / Avg / RMSE |
| 2 | 0.815 / 0.211 / 0.270 | 0.229 / 0.057 / 0.076 | 0.223 / 0.042 / 0.059 |
| 3 | 0.812 / 0.361 / 0.403 | 0.301 / 0.070 / 0.093 | 0.164 / 0.036 / 0.049 |
| 5 | 1.077 / 0.254 / 0.322 | 0.151 / 0.037 / 0.047 | 0.067 / 0.016 / 0.022 |
| 10 | 0.268 / 0.059 / 0.077 | 0.154 / 0.034 / 0.045 | 0.097 / 0.025 / 0.032 |
| 15 | 0.367 / 0.120 / 0.146 | 0.035 / 0.010 / 0.012 | 0.052 / 0.011 / 0.014 |

To follow the desired path, $\psi$ is defined as $\psi\left(C_P \hat{z}_{k+i|k}\right) = \left[X_{k+i|k}, Y_{k+i|k}, \theta_{k+i|k}\right]^T$, and updated in (20):

$$\begin{cases} X_{k+i+1|k} = X_{k+i|k} \\ \quad + T_s \left(V_{x,k+i|k}\cos\theta_{k+i|k} - V_{y,k+i|k}\sin\theta_{k+i|k}\right) \\ Y_{k+i+1|k} = Y_{k+i|k} \\ \quad + T_s \left(V_{x,k+i|k}\sin\theta_{k+i|k} + V_{y,k+i|k}\cos\theta_{k+i|k}\right) \\ \theta_{k+i+1|k} = \theta_{k+i|k} + T_s \omega_{r,k+i|k} \end{cases}$$
(20)

The vector function $h_x$ is defined as:
$$h_x\left(\hat{z}_{k+i|k}\right) = \begin{bmatrix} C_P \hat{z}_{k+i|k} - x_{\max} \\ -C_P \hat{z}_{k+i|k} + x_{\min} \end{bmatrix} \quad (i \in [1, N_p])$$
(21)
to respect the physical constraints of vehicle states, where $x_{\min}$ and $x_{\max}$ are the corresponding minimum and maximum values, respectively.

Moreover, as the decoupling control structure, $T_k$ is generated from the P controller, which remains unchanged during the control of lateral motions. As a result, (15) only finds the solution of wheel steering angle within its physical constraints, and thus $h_u$ is defined as:
$$h_u\left(u_{k+i|k}\right) = \begin{bmatrix} \delta_{f,k+i|k} - \delta_{f\max} \\ -\delta_{f,k+i|k} + \delta_{f\min} \\ T_{k+i|k} - T_k \\ -T_{k+i|k} + T_k \end{bmatrix} \quad (i \in [0, N_c - 1])$$
(22)
where $\delta_{f\min}$ and $\delta_{f\max}$ are the minimum and maximum values of admissible wheel steering angle, respectively. The admissible range is expected to cover the steering maneuvers that allow the vehicle to follow the desired path without exceeding the physical constraints of the steering system. In this study, $\delta_{f\max} = -\delta_{f\min} = 0.3$ rad.

*Remark 3:* Using commercial software – CarSim and a hardware-in-the-loop test platform [5], the collected data were sampled every 0.025 s. In order to facilitate the network's coverage of the feature space, a wide variety of road segments and driving maneuvers are included.

### B. Dimensions of $\Phi_k$

The dimensions of $\Phi_k$ are crucial hyperparameters that need to be specified before building a deep neural network. The higher dimensions strengthen the model fidelity in the lifted space but at the expense of increased training time, storage resources, and the scale of optimization problem. The dimensions of $\Phi_k$ should represent a "trade-off" between all these factors. Therefore, we define "one-step prediction error," i.e., predicting longitudinal velocity, lateral velocity, and yaw rate one step further based on the learned Koopman operator with different dimensions of $\Phi_k$ and comparing with the actual values to evaluate the reconstruction accuracy and select the appropriate dimensions in Fig. 6 and Table I.

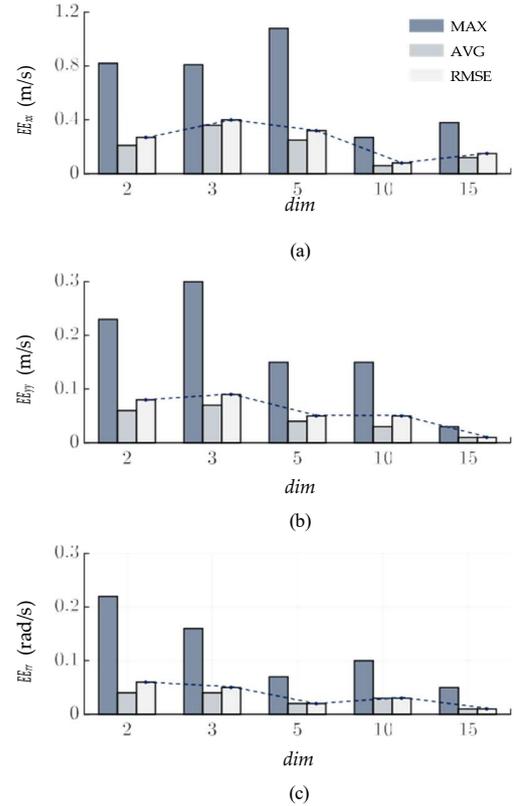

Fig. 6. One-step prediction errors with respect to different dimensions. a) longitudinal velocity. b) lateral velocity. c) yaw rate.

As the number of dimensions increases, the model accuracy incrementally improves in Fig. 6. However, the improvement tends to be saturated beyond the dimensions of 15, followed by a greater network scale. The estimation error for longitudinal velocity of the dimensions of 5 is higher than that of 10 (RMSE: 0.322 m/s (dims=5) > 0.077 m/s (dims=10)), but the estimation errors for lateral velocity are almost identical (RMSE: 0.047 m/s (dims=5), 0.045 m/s (dims=10)). In ad-dition, the 5-dimensional one has less estimation error for the yaw rate (RMSE: 0.022 rad/s (dims=5) < 0.032 rad/s (dims=10)). On the other hand, compared to longitudinal vehicle velocity, control strategies are more sensitive to numerical precision in lateral velocity and yaw rate. In light of the aforementioned considerations, this study suggests dimensions 5 for $\mathbf{\Phi}_k$.

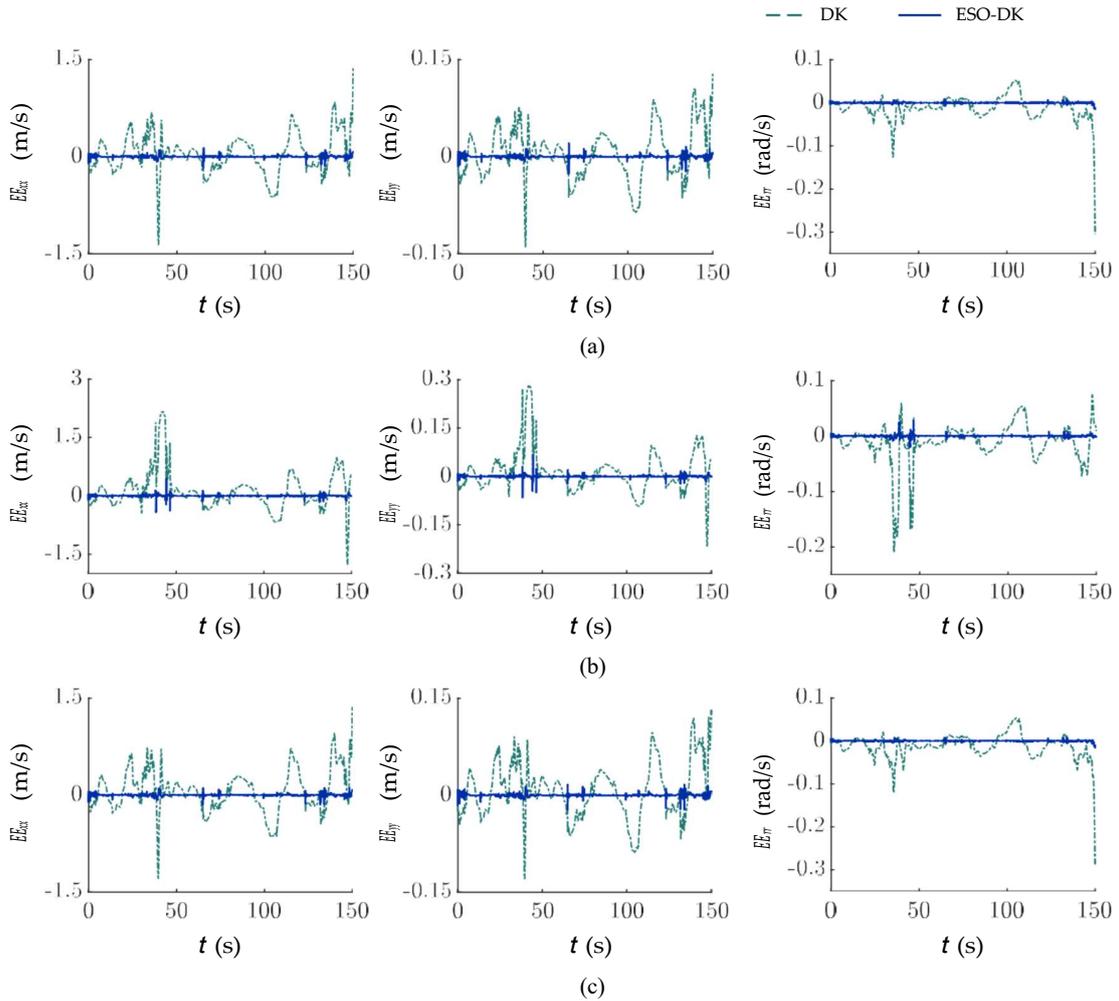

Fig. 7. Comparison results of deep Koopman operator model with ESO (ESO-DK) and without ESO (DK). (a) 1648 kg (sprung mass), 0.85 (road adhesion coefficient). (b) 1648 kg (sprung mass), 0.55 (road adhesion coefficient). (c) 1848 kg (sprung mass), 0.85 (road adhesion coefficient).

*Remark 4:* Once the dimensions of $\mathbf{\Phi}_k$ have been determined, an artificial neural network is built to map the lifting between $\mathbf{x}_k$ and $\mathbf{\Phi}_k$. This network consists of 3 hidden layers, each containing 128 units. Additionally, two single-hidden layers, without activation and bias, are employed to represent $\mathbf{A}$ and $\mathbf{B}$ as introduced in Section II-A. The "Min-Max" method is used to normalize data in order to reduce standard deviations. The steepest descent is chosen for training based on the gradient of the loss function with respect to the network parameters.

## IV. VALIDATION AND RESULTS

In this section, we compare the accuracy and robustness of deep Koopman operator models with and without ESO throughout conditions different from training data. Next, we evaluate the tracking performance and computational efficacy of the ESO-DKMPC on various double-lane change scenarios that induce both linear and nonlinear responses.

### A. Model accuracy and robustness

Yongqian Xiao et al. [27] have discussed and verified the superiority of the Koopman operator model in system modelling over other popular data-driven modelling methods. The DNN-informed Koopman operator model with or without the ESO, denoted as ESO-DK and DK respectively, in Fig. 7.

The initial longitudinal velocity is 40 km/h, the nominal sprung mass of the vehicle is 1648 kg, and the nominal road adhesion coefficient is 0.85 (Fig. 7(a)). Next, we modify the adhesion coefficient to 0.55 (Fig. 7(b)) and the sprung mass to +200 kg over the nominal value (Fig. 7(c)), respectively, to further conduct the comparison.

Overall, the ESO-DK and DK offer promising prospects for modelling and predicting vehicle states under three conditions, but the DK is inferior to the ESO-DK. For the longitudinal velocity, the DK model has notable deviations around 40 s, 100 s, and 150 s, while it exhibits contrary trends in Fig. 7(b). Similar observations can also be found in the results of the lateral velocity and yaw rate. On the other hand, the impact of road adhesion coefficient on the modelling accuracy surpasses that of increased mass, proving its sensitivity to friction changes in vehicle dynamics. Because of the online estimation of total disturbance in the lifted space by ESO, which is a complement to the learned DNN-informed Koopman operator model, the ESO-DK has enhanced model accuracy and robustness.

By incorporating ESO, the feature space is not limited to the collection data but is instead able to extend and generalize real scenarios. Therefore, the ESO-DK model is capable of capturing complex dynamics in real time while maintaining a linear form, thus holding the potential for practical applications.

### B. Double-lane change simulations

To further highlight the proposed ESO-DKMPC for trajectory tracking of autonomous vehicles, four different ways are outlined herein for comparison. Due to the implementation of a decoupling control structure for vehicle trajectory tracking, we emphasize the comparison of the performance of lateral motions. As a result, the proportional control (P control) for the longitudinal profile is utilized by all four methodologies. The primary differences reside in the accuracy of the vehicle dynamics model. Specifically, these distinctions manifest themselves in the following ways:

1) **LMPC**: the linear 2 DoF model (single-track model) is embedded to predict the vehicle states within the prediction horizon [36]. The linear model is widely carried out in real applications due to its simplicity and conciseness for MPC implementation by quadratic programming (QP) techniques, for example, *quadprog* [37].
2) **NMPC**: the nonlinear tire model - Magic Formula (MF), is employed to depict tire dynamics and later obtain the predictive vehicle states to follow the desired trajectory. The nonlinearities of tire forces have been proven effective for model accuracy enhancement but necessitate higher computational resources [38]. The constrained optimization problem is solved using CasADi/IPOPT toolbox [39].
3) **ESO-DKMPC**: the proposed method in this study. The vehicle states are lifted to the high-dimensional state space by the learned deep Koopman operator, and the total disturbances are also estimated using ESO as the complement to improve model fidelity in the lifted space. Based on (16)∼(22), (15) is also solved by CasADi/IPOPT.
4) **DKMPC**: similar to 3). The learned deep Koopman operator model without ESO compensation is adopted to predict vehicle states in the lifted space and generate the control commands.

Moreover, we choose the typical double-lane change test with two different conditions: constant speed on the high-$\mu$ road and varying speed on the low-$\mu$ road, for validation and discussion. In the first scenario, the vehicle travels at a constant and moderate speed, resulting in a smoother steering response, and operates in the linear region. On the other hand, the vehicle operates at a varying and high speed with aggressive steering input under the second scenario, which exhibits nonlinear behaviors. The comparisons enable us further to assess the performances and distinctions among the four methods.

*1) Constant speed on the high-$\mu$ road:* The adhesion coefficient of the road is 0.85, and the longitudinal velocity is maintained at 35 km/h. The comparison results are demonstrated in Fig. 8. In general, all four methods present good tracking performance on high-$\mu$ road in Fig. 8(b), exhibiting the ability to approximate vehicle dynamics in the linear regions. Specifically, the LMPC deviates more when the vehicle crosses the first curve. During the second lane change maneuver, the wheel steering angle of the NMPC (Fig. 8(c)) is greater than that of the LMPC, DKMPC, and ESO-DKMPC, and it fluctuates slightly after returning to the straight, resulting in an evident lateral deviation.

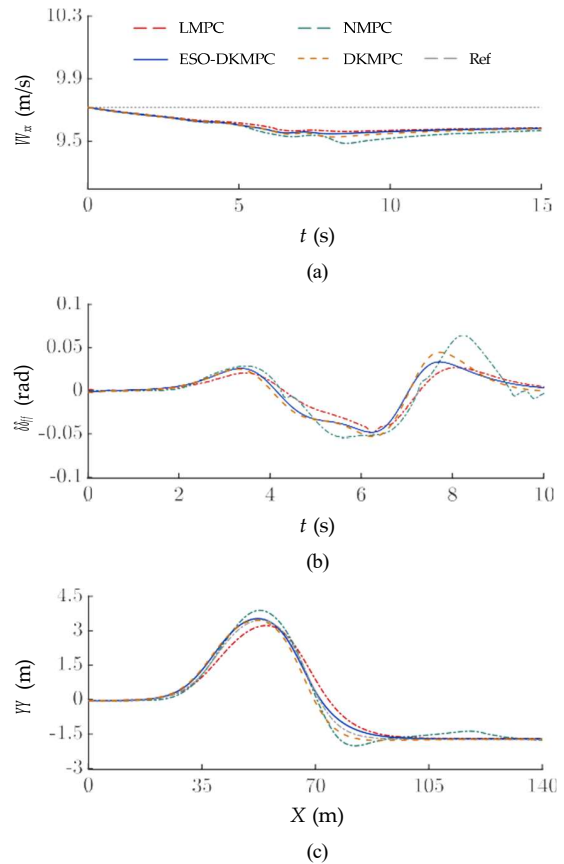

Fig. 8. Simulation results of constant speed on the high-$\mu$ road. (a) Longitudinal velocity. (b) Wheel steering angle. (c) Trajectory.

TABLE II
STATISTIC RESULTS OF CONSTANT SPEED ON THE HIGH-$\mu$ ROAD

|  | $e_Y$ (m) | $\Delta\varphi$ (rad) |
|---|---|---|
|  | Max / Avg / RMSE | Max / Avg / RMSE |
| **LMPC** | 0.606 / 0.136 / 0.225 | 0.104 / 0.025 / 0.039 |
| **NMPC** | 0.619 / 0.167 / 0.229 | 0.106 / 0.025 / 0.034 |
| **DKMPC** | 0.493 / 0.123 / 0.180 | 0.026 / 0.006 / 0.010 |
| **ESO-DKMPC** | 0.217 / 0.066 / 0.093 | 0.042 / 0.012 / 0.017 |

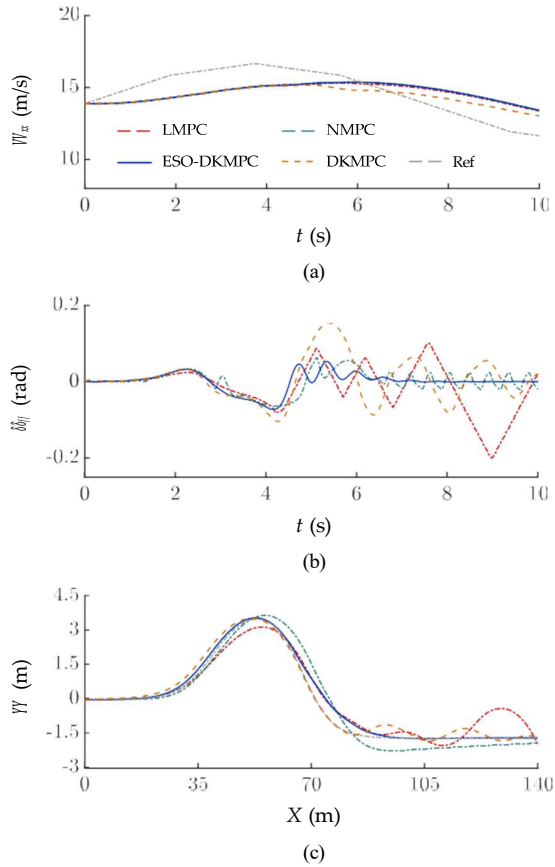

Fig. 9. Simulation results of varying speed on the low-$\mu$ road. (a) Longitudinal velocity. (b) Wheel steering angle. (c) Trajectory.

TABLE III
STATISTIC RESULTS OF VARYING SPEED ON THE LOW-$\mu$ ROAD

|  | $e_Y$ ($m$) | $\Delta\varphi$ ($rad$) |
|---|---|---|
|  | Max / Avg / RMSE | Max / Avg / RMSE |
| **LMPC** | 1.752 / 0.345 / 0.498 | 0.416 / 0.064 / 0.104 |
| **NMPC** | 1.174 / 0.342 / 0.444 | 0.150 / 0.034 / 0.054 |
| **DKMPC** | 0.521 / 0.158 / 0.210 | 0.133 / 0.035 / 0.053 |
| **ESO-DKMPC** | 0.686 / 0.107 / 0.198 | 0.063 / 0.014 / 0.023 |

As a result of the coupling between the longitudinal and lateral dynamics of the vehicle, the lateral motions influence the longitudinal acceleration and, in turn, the longitudinal velocity. Consequently, the longitudinal velocity tracking corresponding to the four methods presents minor variations, but all are less than 0.3 m/s.

Regarding the lateral deviation error $e_Y$ in Table II, the RMSE values of the ESO-DKMPC are 0.093 m. Whereas the RMSE values of $e_Y$ for both the LMPC, NMPC, and DKMPC are greater than 0.180 m. As for the yaw angle error $\Delta\varphi$, the ESO-DKMPC and DKMPC have comparable results with the RSME values less than 0.020 rad. In contrast, the RMSE values of the LMPC and NMPC exceed 0.030 rad. Incorporating with the total disturbance estimation of ESO, the DNN-informed deep Koopman operator model has improved the model fidelity, demonstrating its remarkable potential in representing vehicle dynamics.

*2) Varying speed on the low-$\mu$ road:* The adhesion coefficient of the low-$\mu$ road is 0.5, and the desired longitudinal velocity varies from 45 km/h to 55 km/h with an initial value of 50 km/h (Fig. 9(a)). As the vehicle is expected to follow the trajectory with two curves, and the friction force is limited, both results in the vehicle and tires exhibiting nonlinear characteristics.

In Fig. 9(c), the tracking performance varies on the low-$\mu$ road. The LMPC method returns to the straight road with the most lateral deviation after the second curve. The corresponding wheel steering angle in Fig. 9(b) has an apparent oscillation and tends to diverge, indicating that the vehicle has almost lost stability. This is because the high speed on the low-$\mu$ curved road leads to the tire forces close to the adhesion limit, the linear vehicle dynamics model cannot adequately capture the nonlinear behaviors of the vehicle. Therefore, the LMPC is incapable of controlling the vehicle's lateral motions. As for the NMPC, the nonlinear model has the ability to identify the nonlinear characteristics in such conditions. The vehicle of the NMPC method can keep the stability but causes lateral deviations due to its limited model accuracy. Notably, although the wheel steering angle has a slight fluctuation in Fig. 9(b), the overall trajectory may still be controlled. The DKMPC enhances tracking performance, especially before $X$ = 83 m. Afterward, it exhibits less oscillations in steering angle and lateral deviation compared to the NMPC on the straight road. However, the proposed ESO-DKMPC is superior to the aforementioned three physics-based methods. Because of its high model fidelity to feature nonlinearities, the ESO-DKMPC is capable of stably following the desired path on the low-$\mu$ curved road. Combined with that in Section IV-B1, the results verify the feasibility and robustness of the proposed method. Besides, there are inevitable but acceptable tracking errors when applying the P controller for the longitudinal velocity profile in Fig. 9(a). The errors can be narrowed by calibrating the proportional coefficient, but this is not the focus of this study.

The maximum error of $e_Y$ for the LMPC and NMPC are 1.752 m and 1.174 m in Table III, respectively, while that of the ESO-DKMPC is 0.686 m. Besides, the RSME values of the ESO-DKMPC are also lower than those of the LMPC, NMPC, and DKMPC. In terms of $\Delta\varphi$, the maximum errors of the LMPC, NMPC, and DKMPC are 0.416 rad, 0.150 rad, and 0.133 rad, respectively, while the value for the ESO-DKMPC is 0.063 rad. Similarly, the RSME values of $\Delta\varphi$ for the four methods have the same trends.

When the vehicle speed varies and is high on the low-$\mu$ curved road, the aggressive steering angle enforces the vehicle into the nonlinear region. The linear model, thereby, can hardly represent the nonlinear behaviors. Compared to the linear physics-based approach (LMPC), the nonlinear physics-based (NMPC) significantly elevates the model accuracy. However, the nonlinear physics-based model cannot accurately feature the nonlinearities and consists of a plurality of hyper-parameters that are time-consuming to calibrate and obtain in practical applications. In contrast, the deep learning-informed Koopman operator maps the nonlinearities in the original space to be a linear form in the lifted space by only collecting the input and output data, which proves significant superiority in system modelling. Moreover, the ESO achieves real-time identification of the total disturbance, including unmodelled dynamics and uncertainties in the lifted space, improving model accuracy beyond that of the DKMPC. This renders the proposed ESO-DKMPC feasible and effective under the vehicle working within linear and nonlinear regions.

*C. Computational complexity*

We compare the computation burden of four approaches in Section IV-A and IV-B on *Intel(R) Core(TM) i7-CPU@1.10 GHz*. Given that the network parameters of deep Koopman operator have been finalized by offline training, the recording time does not include the training time. The average computational time: LMPC (2.36 ms) < DKMPC (261 ms) < ESO-DKMPC (370 ms) < NMPC (3076 ms). The computational complexity of NMPC highly increases as a result of its strong nonlinear optimization problem. Due to the linear evolution dynamics in the lifted space, the constrained finite-time optimization of ESO-DKMPC becomes a more computationally efficient problem. While the execution time of the proposed method is more than the LMPC, it is still an acceptable result that can be enhanced for the implementation of the controller. This confirms its applicability in real-world contexts.

## V. CONCLUSIONS

This paper presents a control framework for autonomous vehicles using a deep Koopman operator model incorporating ESO. In terms of trajectory tracking, ESO is designed to online estimate the total disturbances, including modelling errors and residuals, as a complement to the learned DNN-informed Koopman operator model to feature vehicle planar dynamics. The model is then used to formulate the constrained finite-time optimization problem of the model predictive control scheme (ESO-DKMPC) to control the vehicle's lateral motions.

By incorporating ESO, the feature space is not limited to the collection data but is instead able to extend and generalize real scenarios. This demonstrates that the proposed model holds the potential for practical applications. Further, two different conditions under the double-lane change scenario are built on the CarSim/Simulink co-simulation platform. Compared with the LMPC and NMPC informed by the physics-based model, the ESO-DKMPC exhibits superior tracking performance and robustness both within linear and nonlinear regions. The reason for this can be attributed to its higher model fidelity, which is a result of the learned deep Koopman operator and ESO in the lifted space.

Future research will the ESO-DKMPC in more challenging and critical scenarios involving higher speed and changing terrains on the hardware-in-the-loop or on-vehicle platform.

## APPENDIX A
### CONVERGENCY PROOF OF ESO

We define $g_1(z_k) = g_2(z_k) = z_k$, $\tilde{w}_k = w_k - \hat{w}_k$, and assume $w_{k+1} = w_k$. Then, using (12)∼(14), we have:

$$\begin{aligned} z_{k+1} &= (A_\theta + \beta_1 I) z_k + \tilde{w}_k \\ \tilde{w}_{k+1} &= \beta_2 z_k + \tilde{w}_k \end{aligned} \quad (23)$$

Further, the error space is denoted as:

$$\begin{bmatrix} z_{k+1} \\ \tilde{w}_{k+1} \end{bmatrix} = \underbrace{\begin{bmatrix} A_\theta + \beta_1 I & I \\ \beta_2 I & I \end{bmatrix}}_{\Theta} \begin{bmatrix} z_k \\ \tilde{w}_k \end{bmatrix} \quad (24)$$

If the spectral radius of $\Theta$ is less than 1 by choosing appropriate $\beta_1$ and $\beta_2$, i.e., $\rho(\Theta) < 1$, then when $k \to \infty$, both the errors $z_k$ and $\tilde{w}_k$ converge to zero.